\documentstyle[preprint,aps]{revtex}

\begin{document}
\draft
\title{Metal-insulator transition and magnetic ordering in Hubbard models
near the Nagaoka limit}
\author{Q. P. Li and Robert Joynt}
\address{Department of Physics and Applied Superconductivity Center\\
 University of Wisconsin-Madison \\1150 University Avenue,
Madison, Wisconsin 53706 }
\date{\today}
\maketitle
\begin{abstract}
We study the metal-insulator transition and magnetic ordering in
the Hubbard model using the particle-hole mapping.  The analysis simplifies
near the ferromagnetic limit.  We find that the two dimensional(2D)
Hubbard model has a charge excitation gap at half-filling for any
finite U in this region on both the bipartite square lattice and the
nonbipartite triangular lattice. In some cases, the system goes through a
first-order phase transition to become a paramagnetic metal as $S_{z}$ is
lowered. We also discuss the extension to the doped case.  We find that in the
large U limit, a single doped hole has a bandwidth of order of J rather than
t at $S_{z}=0$.
\end{abstract}
\narrowtext

\section{Introduction}

The Hubbard model is one of the most extensively studied strongly
correlated Fermion models in condensed matter
physics.\cite{Hubb,frad,mott,hirs} Despite its simple appearance,  it has
been used to explain a wide variety of physical phenomena such as
ferromagnetism,\cite{nag} antiferromagnetism,\cite{ande}
metal-insulator
transition,\cite{Hubb,mott} and more recently, it has been proposed
as a model for high-$T_{c}$
superconductivity.\cite{azr}  One of the interesting features of the Hubbard
model is the possibility of transforming from the repulsive ($U>0$) to the
attractive ($U<0$) model.  It has been known for a long time
that a positive-U Hubbard model can be mapped
into a negative-U model by making a particle-hole transformation on
one of the spin species.\cite{keme,kris,li}  Here we use this mapping
to study the magnetic and transport  properties of Hubbard
model on various lattices.
One of the motivations to use this transformation is to make
use of the existing knowledge about the attractive
Fermion gas in finding a
good trial wavefunction for the Hubbard model. For example, Leggett\cite{leg}
and Nozieres and
Schmitt-Rink\cite{nozi} have shown that the BCS wavefunction contains the
right physics of attractive Fermion gas in both the weak and the strong
coupling limits, and it may be used
as an interpolation scheme in between to describe the progressive buildup
of pairing correlations in the ground state. Thus we can use a unified
method for all U.  Furthermore, this wavefunction (in the transformed
variables) provides a natural description of the binding and unbinding
of particle-hole-pairs.  This binding can be regarded, in certain situations,
as the Mott transition into the insulating state.

Because of the importance of Hubbard model,
exactly soluble limits or modifications are of great interest, as
evidenced by much work on the Nagaoka limit of small doping and large
U, and on the one-dimensional t-J model. By mapping the positive-U Hubbard
model into a negative-U model, we are able to solve the Hubbard model in
the limit of high total
z spin, $S_{z}$ for all U.  We find that dimensionality plays an important
role in  the metal-insulator transition in this limit. In1D and 2D,
the Hubbard model always has a charge excitation gap at half-filling
at high $S_{z}$, and therefore, is always an insulator for any finite U
in the high-$S_{z}$
region. In three or higher dimensions, however, there exist a critical
interaction $U_{b}$, which separates a metallic phase (small U) from  an
insulating phase (large U).   Using the information from the
exact solution, we construct a BCS type of many-body wavefunction for the
lower $S_{z}$ case and calculate the phase
diagram in the $S_{z}$-U plane. In our work, we choose to fix
$S_{z}$ rather than apply an external magnetic field which acts on the
spins. We discuss this choice in more detail below.

Using a particle-hole transformation for spin-up electrons, the Hubbard
Hamiltonian
\begin{eqnarray}
H = -\sum_{ij\sigma} t_{ij}c_{i\sigma}^{\dagger}c_{j\sigma} + h.c.
+ U\sum_{i} c_{i\uparrow}^{\dagger}c_{i\uparrow}c_{i\downarrow}^{\dagger}
c_{i\downarrow} ,
\end{eqnarray}
can be written as\cite{kris}
\begin{eqnarray}
H = \sum_{ij} t_{ij}h_{i}^{\dagger}h_{j} -
\sum_{ij} t_{ij}d_{i}^{\dagger}d_{j} + h.c. +
 U\sum_{i} d_{i}^{\dagger}d_{i} -
U\sum_{i} d_{i}^{\dagger}d_{i}h_{i}^{\dagger}h_{i} .
\end{eqnarray}

\noindent Here $h_{i}^{\dagger}=c_{i\uparrow}$ and $d_{i}^{\dagger}=
c_{i\downarrow}^{\dagger}$ are the creation operators of the ``hole''
and the ``doublon''. $c_{i\sigma}^{\dagger}$ is the original electron
creation operator. The ``vacuum'' state $|0>$ has an up spin electron at
every site. In this hole-doublon representation (referred as HDR hereafter),
holes and doublons have an on-site attractive interaction $-U$, and
doublons have a site energy $U$. The zero-radius bound state of a doublon
and a hole at site $i$ is nothing but a spin down state at $i$. In the
dilute limit, the bound state pair behaves like a hard-core
boson.\cite{nozi} Also note the change in sign of the hole kinetic
energy, a crucial part of the transformation.

The rest of paper is organized as follows: In Sec. II, we present the
exact solution in the high total spin limit. We show, by calculating the
Kohn charge stiffness
constant, that the system is an insulator when and only when there exists
a bound state solution in HDR. We devote Sec. III to the
lower total spin case. In Sec. IV, we present a possible generalization of
our many-body wavefunction at the half-filling to the hole-doped case and
use it to calculate the minimum energy and the bandwidth of the doped hole.
  We summarize our results in Sec. V.

\section{The high total spin limit}

Let us start with the exactly soluble problem of a single pair of doublon and
hole,
i.e., the band is half-filled and there is only one down spin, $S_{z}=(N-2)/2$.
Here N is the number of electrons. One of the motivation to study the single
pair problem is that in 2D, the existence of the bound state
solution is  the
necessary and sufficient condition of the BCS many-body instability,\cite{rand}
which in our case corresponds to a spin excitation gap at lower $S_{z}$,
although the actual situation is more complicated
because of the particle-hole mapping, as we shall discuss later in Sec. III.

The system consists of a single hole and a single doublon, all other sites
being occupied by an up spin. In this background, both particles move
freely, except for their attraction for one another, which is zero range. This
two particle problem differs from the text book bound state problem only
in that the particles move in a band of finite width and that the hole
prefers to sit at the top, not the bottom, of this band. The latter
circumstance implies that the lowest energy state of a bound pair may not have
zero total momentum.

The wavefunction of the system consisting of one hole and one doublon with
total momentum $\vec{K}$ in the HDR can be written as,
\begin{eqnarray}
\Psi_{\vec{K}} = \sum_{\vec{k}}\alpha_{\vec{k}}
d_{\vec{k}}^{\dagger}h_{\vec{K}-\vec{k}}^{\dagger}|0> .
\end{eqnarray}
Substituting Eqs.(2) and (3) into the Schrodinger equation yields,
\begin{eqnarray}
\alpha_{\vec{k}} = -\frac{AU/N}{E_{\vec{K}}-\epsilon_{\vec{q}}
+\epsilon_{\vec{K}-\vec{q}}} .
\end{eqnarray}

\noindent Here $\epsilon_{k}=-\sum_{j}
t_{ij}e^{i\vec{k}\cdot\vec{r}_{ji}}$ is the
dispersion relation for the non-interaction case and A is a normalization
constant which cancels out in the eigenvalue equation. The eigenenergy of the
system  is given by
\begin{eqnarray}
\frac{1}{N} \sum_{\vec{q}} \frac{1}{E_{\vec{K}}-\epsilon_{\vec{q}}
+\epsilon_{\vec{K}-\vec{q}}} =
-\frac{1}{U} .
\end{eqnarray}

 We first consider
the most simple case that
$t_{ij}$ is only nonzero for nearest neighbor hopping on a
d-dimensional hyper-cubic lattice, then $\epsilon_{k}=-2t\sum_{m=1}^{d} cos
k_{m}$. (We assume the lattice constant $a=1$ throughout this paper.) The
doublon kinetic energy is minimum at $\vec{k} = 0$, and the hole kinetic
energy at $\vec{k}=(\pi,\pi,...,\pi)$.
For every given $\vec{K}$, one can calculate $E_{\vec{K}}$ from Eq.(5).
We find that the minimum value of
$E_{K}$ is reached at $K_{1}=K_{2}=\ldots=K_{d}=\pi$, consistent with
the kinetic energies of the particles.  So
Eq. (5) can be written as
\begin{eqnarray}
f(E) = \int \frac{d^{d}q}{(2\pi)^{d}}
\frac{1}{E+4t\sum_{m=1}^{d} cos q_{m}} = -\frac{1}{U}
\end{eqnarray}
in the limit $N \rightarrow \infty$. Here E is the minimum value of
$E_{\vec{K}}$ and the integration is
over the first Brillouin zone. The structure of the solutions
of Eq. (6) is well known.\cite{schr:sc} There is an extended state
continuum of eigenvalue E between
energy $-4td$ and $4td$, and there may or may not be a bound state below
the continuum lower edge $E_{-}=-4td$ depending on the interaction U and
the dimensionality d. For d=1, the integral in Eq. (6) is divergent as
$1/\epsilon$ when E approaches $E_{-}$ from below so that
$f(E\rightarrow E_{-}) \rightarrow -\infty$. Also
$f(E\rightarrow -\infty) = 0$. This means that
there is always a bound state solution no matter how small the
interaction U is. $d=2$ is the marginal case, the integral in Eq. (6)
is logarithmically divergent as E approaches $E_{-}$ from below.
So there is a bound state for any U, although the gap
between the bound state and the extended state continuum decreases
exponentially as $e^{-8\pi t/U}$ when U is small. For d=3, the
integral of Eq. (6) is convergent as E approaches $E_{-}=-12t$ and
there exists a critical $U_{b}$ such that when $U<U_{b}$,
there is no bound state.  $U_{b}$ is given by
\begin{eqnarray}
\frac{1}{U_{b}} = \int \frac{d^{3}q}{(2\pi)^{3}}
\frac{1}{4t(3 - \sum_{m=1}^{3} \cos q_{m} )} .
\end{eqnarray}
Numerical evaluation of the integral gives $U_{b}\simeq 7.916t$. $U_{b}$
increases monotonically as d is increased past 3.

Next we consider the single-pair hole-doublon problem on a 2D triangular
lattice with the
nearest neighbor hopping only. The sign
difference of the hopping terms of doublons and holes due to the
particle-hole transformation now becomes important. When the system is on
a bipartite lattice,  the minus sign can be simply ``gauged'' away by
redefining $h_{i}^{\dagger}=s(i)c_{i\uparrow}$. Here s(i) equals 1 or -1
when the site i belongs to two different sublattices. After this
redefinition, the new ``hole'' has the same hopping term as doublons
and is able to hop coherently with the doublon which it forms a
bound state with. Now the minimum of energy is reached when the transformed
total momentum $\vec{K}=0$. When the Hubbard model is on a nonbipartite
triangular lattice, however, the sign difference of the hopping terms of
holes and doublons in Eq. (2) can not be gauged away, i.e., the hopping
terms with $+t$ and $-t$ are not symmetric any more. For example, the
minimum value of the doublon hopping term is $-6t$ on a 2D triangular lattice,
while the minimum
value of the hole hopping term is only $-3t$. So the lower edge of the
extended state continuum is at $E_{-}=-9t$ instead of $-2zt=-12t$.
(Here z=6 is the number of the nearest neighbors.)  The eigenenergy of a
single pair of hole and doublon on a triangular lattice is still given by
Eq. (5). Now
\begin{eqnarray}
\epsilon_{\vec{k}}=-2t[cos k_{1} + cos k_{2} + cos(k_{1}-k_{2})].
\end{eqnarray}

\noindent Here we have chosen the two primitive axes of the triangular
lattice as the axes
of our coordinate system. $k_{1}$ and $k_{2}$ are the components of
$\vec{k}$ in the non-orthogonal basis which generates the reciprocal lattice.
The minimum of $E_{\vec{K}}$ is
reached when the total momentum of the hole-doublon pair
$\vec{K}$ is at one of the corners of the hexagonal Brillouin zone, for
example, $\vec{K}=(2\pi/3,-2\pi/3)$ in the non-orthogonal basis. Then Eq. (5)
becomes
\begin{eqnarray}
A_{c}\int \frac{d^{2}q}{(2\pi)^{2}} \frac{1}{E+2\sqrt{3}t [sin
(\frac{\pi}{3}-q_{1}) + sin (\frac{\pi}{3}+q_{2}) + sin (\frac{2\pi}{3}-
q_{1}+q_{2})]} = -\frac{1}{U} .
\end{eqnarray}

\noindent Here $A_{c}=\sqrt{3}/2$ is the area of the primitive cell and the
integral is over the first Brillouin zone. The integral of Eq. (9) is
logarithmically divergent as E
approaches the lower edge of the extended state continuum $E_{-}=-9t$
from below, which means that, just as on the square lattice, a single
pair of hole and
doublon on the nonbipartite triangular lattice also always forms a bound
state no matter how small the interaction U is. Similarly, we can show that
the conclusion is also valid for the 2D Hubbard model with the nearest
and the next nearest neighbor hopping. In this case, the noninteracting
energy dispersion on a square lattice is given by
\begin{eqnarray}
\epsilon_{k}=-2t[cos k_{x} + cos k_{x}] - 4t_{2} cos k_{x} cos k_{y}.
\end{eqnarray}

\noindent Here the first and the second term correspond to the nearest
neighbor and the next nearest neighbor hopping, respectively. For
$t>2t_{2}>0$, the minimum of $E_{K}$ is at $\vec{K} = (\pi,\pi)$, which is
the same as the $t_{2}=0$ case.
We can see that the question of whether a pair of hole and
doublon always form a bound state for any finite U is only determined
by the dimensionality of the
system. For $d>2$, there exists a critical $U_{b}$, when and only when
$U>U_{b}$, doublons and holes form charge neutral bound states; While
for $d\leq 2$, they always form bound states for any finite U. This conclusion
is easily proved by considering the behavior of the integral in Eq.(5) near
the point in k-space where $\epsilon_{\vec{k}} -\epsilon_{\vec{K}-\vec{k}}$
takes on its minimum value $E_{-}$. This is a variant of standard arguments
about bound states. Its interest here is that two dimension is more similar
to one dimension than it is to three dimensions in this limit of the
Hubbard model, namely for high total spin but for {\em all} U. This is
contrary to the zero total spin case, where it is usually felt that one
dimension is very special as regards the metal-insulator transition.

The conclusion that a bound state exists for all U in one and two
dimensions and above a critical U in three dimension is independent
of lattice structures. This is well known in single particle quantum
mechanics. We have simply reinterpreted the result as a metal-insulator
transition at high $S_{z}$. The critical dimension for this transition
is two in this model.

To verify that the bound state is indeed an insulating state,
we follow Kohn\cite{kohn} and calculate the charge stiffness constant $D_{c}$
which measures the response of the system to an electromagnetic field.
We assume periodic boundary conditions along the x-direction and
thread the system with a flux $\Phi$, which we represent by a vector
potential $\vec{A}=(\Phi/N_{x})\hat{e_{x}}$, where $N_{x}$ is the
number of sites in the x-direction. The charge stiffness constant
$D_{c}=N_{x}^{2}d^{2}E(\Phi)/d\Phi^{2}$ is a measure of how good
a conductor the system is, and vanishes for an
insulator.\cite{kohn,mill,sark} Here $E(\Phi)$, the ground state
energy in the presence of the flux $\Phi$, is given by,
\begin{eqnarray}
\frac{1}{N} \sum_{\vec{q}} \frac{1}{E(\Phi)-\epsilon_{\vec{q}}(\Phi)+
 \epsilon_{\vec{K}-\vec{q}}(-\Phi)} =
-\frac{1}{U}. \label{dc}
\end{eqnarray}
\noindent Here $\epsilon_{\vec{q}}(\Phi)$ is the single particle kinetic
energy in the presence of the
flux $\Phi$. For a 2D square lattice with the nearest neighbor hopping, for
example, $\epsilon_{\vec{q}}(\Phi)=-2t[cos(q_{x}+\Phi/N_{x}) +
cos(q_{y})]$.  One
can easily write down similar expressions for other cases. We expand
Eq. (\ref{dc})
up to the second order of $\Phi$.  The nth order equation gives
$d^{n}E/d\Phi^{n}$ at $\Phi = 0$. After some algebra, one finds that
$dE/d\Phi$  is always zero, and $d^{2}E/d\Phi^{2}$ is zero
when and only when $E<E_{-}$, i.e., $D_{c}=0$ when and only when there
exists a bound state below the
extended state band. This is because that only when $E<E_{-}=-8t$, one can
expand Eq. (\ref{dc}) in terms of $\Phi/[E+4t(cos q_{x} + cos q_{y})]$;
When $E \geq E_{-}$, this expansion is not valid in some regions of the
momentum
space and $D_{c}$ is finite. This conclusion also applies to the 3D cubic
lattice and other cases. We have therefore established the one-to-one
correspondence between the existence of the bound state and the insulating
behavior of the system.  The bound state wavefunction contains precisely
the hole-doublon phase coherence necessary for insulating behavior.
This coherence can not be achieved when hole-doublon binding is
accomplished by Gutzwiller-type projection methods.\cite{mill}

Next we consider an anisotropic three-dimensional system with small
interlayer hopping.  This is
important in the study of high-$T_{c}$ superconductors. As we
discussed above, the dimensionality of the system plays a very
important role in determining whether the system is  a Mott insulator
or a metal. We can expect that even a small interlayer hopping could be
quite important because it introduces the third dimension.
Mathematically, the interlayer hopping term provides a cutoff to the 2D
logarithmic divergence of Eq. (5).  In this case, Eq. (5) can be written as
\begin{eqnarray}
\int \frac{d^{3}q}{(2\pi)^{3}} \frac{1}{E+4t(cos q_{x}+cos q_{y}) +
4t^{\prime} cos q_{z}} = -\frac{1}{U} .
\end{eqnarray}

\noindent Here $4t^{\prime} cos q_{z}$ is the interlayer hopping term.
As before for $t^{\prime} \neq 0$, there exists a
$U_{b}(t^{\prime})$ such that when and only when $U>U_{b}(t^{\prime})$,
Eq. (11) has a bound state solution. In Fig. 1, we show the critical
interaction
$U_{b}(t^{\prime})$ as a function of $t^{\prime}$. We can see that
$U_{b}$ does depend on $t^{\prime}$ quite strongly in the small $t^{\prime}$
region. For example, at
$t^{\prime}=0.001t$, we still have $U_{b}=2.434t$ even though $U_{b}=0$ at
$t^{\prime}=0$. The crossover to 3D behavior occurs very rapidly as
$t^{\prime}$ is turned on.

When there are few doublons and holes, i.e., in the dilute limit, the
overlap or interaction between the
hole-doublon bound states (HDBS) is small and negligible, and the
single pair HDBS solution gives a correct physical picture of the
system. In
1D and 2D, the Hubbard model always has a charge gap at half-filling
at high $S_{z}$, and therefore, is always an insulator for any finite U
in the high-$S_{z}$
regime. (At least at $S_{z}=N/2$, which is obvious, and at $S_{z}=N/2-1$,
which we have just proved rigorously). In 3D, when the interaction U is
smaller than $U_{b}$, there is no bound state and the system is a
paramagnetic metal (or
semimetal since the carrier density is small); when U is larger than
$U_{b}$, the doublons and holes form charge neutral bound states and
the ground state is a gas of bound doublon-hole pairs,
a very appealing model for  an insulator.\cite{Hubb,kohn,mill}

As the number of HDBS increases,
the bound states merge into an energy band which corresponds to the
lower Hubbard band in the original electron representation.
Similarly, the
extended state continuum corresponds to the upper Hubbard band, and
the gap between the bound state and the extended state is
nothing but the Mott-Hubbard gap.
In 2D, we have shown that a single pair of hole and doublon
always form a bound state and has a gap no matter how small the interaction
U is. The question is whether this gap will survive the overlap and
the interaction between the HDBS pairs. That is the question we shall
address in the next section.

\section{General total spin}

In this section we consider the Hubbard model at half-filling with
general $S_{z}=(N-2M)/2$, i.e., there are M spin-down electrons (or M
doublon-hole pairs in HDR) in the half-filling N-site system.
We exploit the fact that the relationship between the high spin limit and
the lower spin
case is very similar to that between the Cooper pairing
problem and the BCS many-body problem. It is therefore natural for us
to choose a BCS type of
many-body wavefunction for the many-pair state.\cite{schr:sc,fuld}
\begin{equation}
|\vec{K}, S_{z}> = \prod_{k} (u_{k}+v_{k}d_{k}^{\dagger}h_{K-k}^{\dagger})|0>
\label{bcs}
\end{equation}
\noindent where $u_{k}^{2} = \frac{1}{2}(1+\xi_{k}/E_{k})$, $v_{k}^{2}
= \frac{1}{2}(1-\xi_{k}/E_{k})$, $E_{k}^{2} = 4\Delta^{2} +
\xi_{k}^{2}$, and $\xi_{k}=\epsilon_{k}-\epsilon_{K-k}-\mu_{d}-\mu_{h}$
is the kinetic energy of the Cooper pair measured from the Fermi surface.
$\mu_{d}$ and $\mu_{h}$ are the
chemical potential of the doublons and the holes, respectively, and
$\Delta=\frac{U}{N} \sum_{\vec{k}}u_{k}v_{k}$ is the BCS gap function which is
momentum-independent in this case because of the short-range interaction
in Hubbard model.
$\vec{K}$ is the total momentum of the hole-doublon ``Cooper
pairs'' which can also be thought as the wavevector of the
spin-density wave in the original electronic picture.\cite{kris,fuld} This
kind of pairing state was first studied by Fulde and Ferrell, and Larkin and
Ovchinnikov in the
context of superconductivity.\cite{fuld,lark} Obviously this wavefunction
is not an eigenfunction of operator $\hat{S}_{z}$, and the labeling $S_{z}$
on the left-hand
side of Eq.(\ref{bcs}) should be understood  in the sense of the average
expectation
value of operator $\hat{S}_{z}$ just as the number of particles in the
BCS case.
Usually $\vec{K}$ is at one of the corners of the first Brillouin zone as we
can see from the single-pair solution. It could
move away from the corners in some cases, which corresponds to the
incommensurate spin-density wave.

 We evaluate the expectation value of the Hamiltonian Eq. (2)
using the wavefunction of  (\ref{bcs}). The gap equation is then given by
minimizing the energy expectation value\cite{dege:sc}
\begin{eqnarray}
\frac{1}{N} \sum_{\vec{k}}
\frac{1}{[(\epsilon_{k}-\epsilon_{K-k}-2\mu)^{2}+
4\Delta^{2}]^{1/2}} = \frac{1}{U} .
\label{gap}
\end{eqnarray}
\noindent Here $2\mu=\mu_{d}+\mu_{h}$.  The total z spin is given by
$S_{z} = N/2 - \sum_{k}
v_{k}^{2}$, which, together with the gap equation, can be used to calculate the
chemical potential $\mu$ and the gap $\Delta$,
\begin{eqnarray}
\frac{1}{N} \sum_{\vec{k}}
[1 - \frac{\epsilon_{k}-\epsilon_{K-k}-2\mu}{[(\epsilon_{k}-\epsilon_{K-k}-
2\mu)^{2}+4\Delta^{2}]^{1/2}}] = 1 - \frac{2S_{z}}{N} .
\label{chem}
\end{eqnarray}

\noindent Here N is the number of sites. It is easy to demonstrate
that in 2D, Eq.(\ref{gap}) always has a nonzero solution $\Delta$ for any
finite U, which means that
there is always a charge excitation gap in this particular wavefunction
for any finite value of U. To show
that, we take the limit $N \rightarrow \infty$ and replace the sum
$\sum_{\vec{k}}$ with
the integral over the first Brillouin zone,
\begin{eqnarray}
A_{c}\int \frac{d^{2}k}{(2\pi)^{2}}
\frac{1}{[(\epsilon_{k}-\epsilon_{K-k}-2\mu)^{2}+
4\Delta^{2}]^{1/2}} = \frac{1}{U} .
\label{gap2}
\end{eqnarray}
\noindent Here $A_{c}$ is the area of the primitive unit cell.
$A_{c}=1$ for the square lattice and $A_{c}=\sqrt{3}/2$ for the triangular
lattice. The integral in Eq. (\ref{gap2})
is logarithmically divergent as $\Delta$ goes to zero.  This is due to the
fact that generally in $d=2$,
($\epsilon_{\vec{k}}-\epsilon_{\vec{K-k}}-2\mu$) vanishes along a curve in
the momentum
space. That is sufficient to cause the logarithmic divergence of the
left-hand-side of Eq. (\ref{gap2}) as $\Delta\rightarrow 0$.

This reasoning would suggest that the 2D Hubbard model
is an insulator at half filling for any finite U, as is true for the
1D model, and the square lattice with the nearest neighbor hopping only.
To understand how the reasoning breaks down,
 we consider  three cases: \\
(A). bipartite square
lattice with the nearest neighbor (NN) hopping only, \\
(B). square lattice with NN and the next nearest neighbor (NNN)
hopping, and \\
(C). nonbipartite triangular lattice with nearest neighbor hopping only.

\noindent{\bf A. Square lattice with NN hopping only}

In this case, $\epsilon_{\vec{K}-\vec{k}} = -\epsilon_{\vec{k}}$, and the
many-body
wavefunction  (\ref{bcs}) at  $\Delta=0$ is identical to the noninteracting
wavefunction. The fact that the gap equation always has a nonzero solution
for any finite U means that the Fermi liquid state is unstable against
  the BCS state (\ref{bcs}) at finite U.
The staggered magnetization of Hubbard model is proportional to $\Delta/U$.
So obviously a nonzero $\Delta$ means that the system is antiferromagnetically
ordered.  We therefore conclude that the system
is always an antiferromagnetic insulator for any finite U. This result is
well known, but the present method gives a new way of visualizing it.
Consider the first Brillouin zone in Fig.2(a). The doublons occupy the
region near $\vec{K}=0$, and the holes occupy the region near
$\vec{K}=(\pi, \pi)$. The energy curves are exact translates
of one another through $(\pi, \pi)$  for {\em all} fillings, so each
hole pairs with one doublon. This particular model is known to have
perfect nesting at half filling, but note that this property is even
stronger.  There is a perfect particle-hole nesting at all values
of the magnetization.  As a result, the pairing state is always energetically
favorable.

In the large U limit, one can solve the gap equation (\ref{gap}) by expanding
the gap function $\Delta$ in terms of $t/U$. At $S_{z}=0$, we find
\begin{equation}
\Delta = \frac{1}{2} U[1-8\frac{t^{2}}{U^{2}}+O(\frac{t^{4}}{U^{4}})] ,
\label{taylor}
\end{equation}
and the expectation value of the Hamiltonian in the pairing state is
\begin{eqnarray}
E_{0} = <\vec{K},S_{z}=0|H|\vec{K},S_{z}=0> =
-4Nt[\frac{t}{U}+O(\frac{t^{3}}{U^{3}})] ,
\end{eqnarray}
which is the expected result $-NJ$ for a N\'{e}el state, and shows that our
mean-field solution has the correct large-U limit. Here $J=4t^{2}/U$.
The pairing state (\ref{bcs})
may do better when U is small. The reason is that in this case the size of
bound states becomes large, there are many particles inside the region
covered by a single bound state wavefunction,  which will suppress
the relative fluctuation in the interaction effects, and the BCS
mean-field wavefunction becomes a better approximation to the exact ground
state.

\noindent{\bf B. Square lattice with NN and NNN hopping}

The noninteracting energy dispersion in this case is given by Eq. (10).
The crucial difference between the case (A) and (B) is that for case (B),
$\epsilon_{K-k} \neq -\epsilon_{k}$, and the BCS many-body
wavefunction (\ref{bcs}) does not reduce to the noninteracting wavefunction
when $\Delta=0$, i.e., the fact that the gap equation (\ref{gap}) always
has a nonzero
solution for any finite U just means that the wavefunction (\ref{bcs}) at
$\Delta = 0$ is always unstable against the finite-$\Delta$ BCS pairing state,
but the $\Delta = 0$ wavefunction is {\em not} necessarily the Fermi liquid
wavefunction.
 Thus in HDR, when the holes and doublons take advantage of
the attractive interaction between them and form the BCS condensate, they pay
the price of having a little higher kinetic energy in case (B). This can be
seen in Fig. 2(b), where the doublon Fermi surface near ${\bf K}=(0,0)$ can
{\em not} be mapped onto the hole Fermi surface near ${\bf K}=(\pi,\pi)$
through a linear transformation ${\bf k}^{\prime} = {\bf K} - {\bf k}$.
It is easy to show that the Fermi liquid
state has lower kinetic energy than the pairing state (\ref{bcs}). This
problem becomes severe at low $S_{z}$.  So when the gain from
the condensation energy is not enough to compensate the extra kinetic energy,
the BCS state (\ref{bcs}) becomes
unstable to the Fermi liquid state, the system goes through a first order
phase transition and becomes a
paramagnetic metal (Fermi liquid). Once we have established the correct
physical picture, it becomes straightforward to calculate the metal-insulator
phase boundary on which the energy of the many-body wavefunction (\ref{bcs})
is equal to the energy of a true Fermi liquid state. In Fig. 3, we show the
calculated phase diagram of the case (B) with
$t_{2}=0.25t$ in the $S_{z}$-U plane. For $S_{z}=0$, the critical interaction
$U_{c}$ is about 2.3t, which is in agreement with the result of the
numerical and Hartree-Fock calculation by Lin and Hirsch.\cite{hirs} As $t_{2}$
goes to zero, case (B) reduces to case (A), and the critical interaction of the
metal-insulator transition $U_{c}$ goes to zero.

\noindent{\bf C. Triangular lattice with NN hopping}

Case (C) is  more complicated than case (B),  due to the conflict
between  the ``pairing'' property of the wavefunction (\ref{bcs}) and the
``tripartite'' character
of the lattice. As a result, the holes of the wavefunction (\ref{bcs}) only
occupy
three of the six corners of the hexagonal first Brillouin zone (BZ),
leaving the low energy states of the other three corners unoccupied or
partially occupied.
The occupied three corners, as well as the unoccupied three corners, are
related by the reciprocal lattice vectors and form two subsets, respectively,
 but these two
subsets are {\em not} related by the reciprocal lattice vectors. As shown in
Fig.4(a), the hexagonal BZ (dotted line) is equivalent to the parallelogram
(solid line) which is easier to deal with analytically. So by choosing
${\bf K}$ to be at one of the BZ corners (or equivalently, two points A and B
on the diagonal of the parallelogram which divided the diagonal into three
equal segments), we force the system breaking
the point group symmetry as well as the time reversal symmetry. In Fig. 4(a),
we also show the Fermi surfaces of the doublons (solid line) and holes
(dashed line) at $S_{z}=0.4N$ and $0.3N$, respectively. Apparently there are
two minimum energy positions (which correspond to two distinguish sets of
corners of the hexagonal BZ) for holes to occupy, but in the pairing state
(\ref{bcs}) they can only occupy one of them and leave the other unoccupied.
This, as well as the distortion of the Fermi
surface, costs extra kinetic energy, which opens doors for those
many-body states other than the Fermi-liquid or the state of
(\ref{bcs}) with {\bf K} at the one of the corners of BZ as the possible
candidate of the ground state in some parameter
region. One possible candidate is the incommensurate spin-density wave state
which corresponds to a Fulde-Ferrell-Larkin-Ovchinnikov (FFLO) type of state
in HDR. In comparison, the four
corners of the BZ of the square lattice are
related by the reciprocal lattice vector and the increasing of the kinetic
energy in the pairing state in the NNN hoping case is solely due to the
distortion of the Fermi surface. For this reason, we expect the critical
interaction $U_{c}$ of the instability of the BCS state (\ref{bcs}) in case
(C) to be  larger than that in
case (B), and the system may go through other kind of states before becoming a
paramagnetic metal (Fermi liquid).

We  calculate the energy of the pairing state (\ref{bcs}) for general
${\bf K}$.  We find that the energy of the pairing states  ${\bf K}$
at the corner of first Brillouin zone is at least a local minimum. It is
quite possibly also a global minimum. Since there are two set of the
optimum ${\bf K}$ which are not related to each other by the reciprocal
lattice vectors,  the triangular lattice has two degenerate ground
states which may have some interesting consequences. For example, the system
may be vulnerable to the Jahn-Teller distortion.

In Fig. 4(b),  we show the critical interaction $U_{c}$ of the Hubbard model on
a triangular lattice as a function of
total spin $S_{z}$. Below $U_{c}$, the BCS state (\ref{bcs}) is unstable
against the paramagnetic Fermi liquid state. We find, as expected, that
$U_{c}$ is larger than the square lattice with NNN hoping case due to
the incompatibility of ``pairing'' and the tripartite nature of the lattice.
Our calculated critical interaction of the 2D Hubbard model on a
triangular lattice $U_{c}=4.936t$ at $S_{z}=0$  is a little bit smaller than
the result of Krishnamurthy et. al. \cite{kris,sark}, their metal-insulator
transition is at $U_{c2}=5.27t$. But our result is not in
contradiction with their result, because Krishnamurthy et al. have
considered a more complicated metallic state, namely the spin density
wave metallic state while we only compare our pairing state with the
most simple Fermi liquid state. So our result shown in Fig. 4(b) is the lower
bound of the critical interaction of the metal-insulator transition.

\section{Away from half-filling}

In section II and III, we discussed the properties of Hubbard model on various
lattices at half-filling. In HDR, this corresponds to the case where the number
 of doublons is equal to the number of holes. In this section we consider
the case of Hubbard model away from the half-filling. One
natural choice for the wavefunction of the hole-doped system is
\begin{eqnarray}
|\psi_{n}> = \sum_{\{q_{i}\}} A(\{q_{i}\})\prod_{i=1}^{n}
\gamma_{q_{i}}^{\dagger}|\psi_{0}>,
\label{dope}
\end{eqnarray}

\noindent where $\gamma_{p}^{\dagger}=u_{p}h_{K-p}^{\dagger} +
v_{p}d_{p}$ is the creation operator of one of the Bogoliubov quasiparticle
species,\cite{dege:sc,bogo} $|\psi_{0}> \equiv |{\bf K}, S_{z}>$ is the
pairing wavefunction for the half-filled case (cf. Eq. (\ref{bcs})).
$n=N_{h}-N_{d}$, $N_{h}$ and $N_{d}$ is the
number of holes and doublons, respectively. The wavefunction (\ref{dope})
should remain reasonably good for the multi-hole-doped system as long as it
still has the long-range order. Especially, the wavefunction of
the system with one extra-hole (n=1) is
\begin{equation}
|\psi_{1}> = \sum_{q} A_{q}
\gamma_{q}^{\dagger}|\psi_{0}>,
\label{dope1}
\end{equation}

\noindent which represents a series of states from the Nagaoka state to the
antiferromagnetic state with one extra-hole as one moves from high-$S_{z}$ to
$S_{z}=0$.
We evaluate the expectation value of $H$
in the state (\ref{dope1}),
\begin{eqnarray}
<\psi_{1}|H|\psi_{1}> = \sum_{\vec{k}}(\epsilon_{k} -
\epsilon_{K-k})v_{k}^{2}
 -N\frac{\Delta^{2}}{U} + N_{d}U(1-\frac{N_{d}}{N}) + E_{1}(\{A_{k}\}) .
\end{eqnarray}

\noindent Here $E_{1}(\{A_{k}\})=<\psi_{1}|H|\psi_{1}> - <\psi_{0}|H|\psi_{0}>$
is the energy of the extra-doped hole,
\begin{eqnarray}
E_{1}(\{A_{k}\}) = \sum_{\vec{k}}[2\Delta u_{k}v_{k} -
U(1-2\frac{N_{d}}{N})v_{k}^{2} - \epsilon_{k}v_{k}^{2} -
\epsilon_{K-k}u_{k}^{2}]|A_{k}|^{2} - U\frac{N_{d}}{N}.
\end{eqnarray}

\noindent From that we can find the optimum $\{A_{k}\}$ which minimizes
$E_{1}$ with the constraint $\sum_{\vec{k}}|A_{k}|^{2} = 1$ as well as
the bandwidth of the doped hole which is the difference between the
maximum and the minimum of $E_{1}$.  For simplicity,
we consider the 2D Hubbard model on a square lattice with
the nearest neighbor hopping only.   We find that the optimum
$A_{k}=\delta(\epsilon_{k}-\mu-\eta)$ and
the minimum energy of the doped hole is
\begin{eqnarray}
E_{min} = \sqrt{\eta^{2} + \Delta^{2}} - U/2 +
(\mu+\frac{U}{2}-U\frac{N_{d}}{N})\eta/\sqrt{\eta^{2} + \Delta^{2}}.
\end{eqnarray}
\noindent Here $\eta$ is the optimum $(\epsilon_{k}-\mu)$
which minimizes the hole energy. It is given by
\begin{equation}
\eta^{3}+\eta\Delta^{2}+(\mu+\frac{U}{2}-U\frac{N_{d}}{N})\Delta^{2} = 0,
\label{opti}
\end{equation}
with the constraint
\begin{equation}
-4t-\mu \leq \eta \leq 4t-\mu .
\label{cons}
\end{equation}

 When $S_{z}$ is zero, $N_{d}/N=1/2$,
$\mu = 0$,  therefore $\eta = 0$, i.e., the hole is right on the
fermi surface, and
\begin{equation}
E_{min} = \Delta - U/2.
\label{eh}
\end{equation}
Here $\Delta$ is the solution
of the gap equation (\ref{gap}). In Fig. 5(a), we show the minimum hole
energy $E_{min}$
as a function of U/t for a square lattice with the nearest-neighbor hopping
only. One can see that in the small U region, $E_{min}$ decreases
as U increases. It reaches a minimum at $U/t \approx 2.7$. In the large U
limit, $\Delta$ can be written down explicitly as a Taylor series of
$t/U$ (cf. Eq. (\ref{taylor})). We get
\begin{eqnarray}
E_{min} = -\frac{4t^{2}}{U} + O(\frac{t^{4}}{U^{3}})
= -J + O(\frac{t^{4}}{U^{3}}).
\end{eqnarray}

Similarly, we find that the maximum of the hole energy at $S_{z}=0$ is,
\begin{eqnarray}
E_{max} = \sqrt{(4t)^{2} + \Delta^{2}} - U/2.
\end{eqnarray}
In the large U limit,
\begin{eqnarray}
E_{max} = 12\frac{t^{2}}{U} + O(\frac{t^{4}}{U^{3}})
= 3J + O(\frac{t^{4}}{U^{3}}),
\end{eqnarray}
and the bandwidth of the doped hole,
\begin{equation}
W= 4J + O(\frac{t^{4}}{U^{3}}).
\end{equation}
This is in agreement with the recent numerical calculations which find that
a hole in an antiferromagnetic background has a bandwidth in the scale of $J$
rather than $t$, in the small $J/t$ limit.\cite{dag,von,trug,poil,chub}

In Fig. 5(b), we plot the bandwidth
\begin{equation}
W= E_{max} - E_{min} = \sqrt{(4t)^{2} + \Delta^{2}} - \Delta
\end{equation}
as a function of U at $S_{z}=0$. We can see that W is proportional to t
in the small U region where $\Delta \ll t$. It smoothly crosses over to
scale as J in the large U region where $\Delta \gg t$. Physically, this is
in agreement with the string picture of a hole in an antiferromagnetic
environment.\cite{trug,dag2} Due to the presence of the frustrated spin
string left behind
the hopping hole, the hole acquires a large effective mass $m^{*}$ which is
proportional to the inverse of the effective hopping $t^{*}$, and therefore,
also to the inverse of the bandwidth W. Due to the presence of the
interaction, $t^{*}$ is renormalized
from t to $J=4t^{2}/U$ in the large U limit.

When $S_{z} \neq 0$, the chemical potential $\mu$ is not zero anymore and is
given by Eq. (\ref{chem}).  In Fig. 6 we show the minimum hole energy as a
function of $S_{z}$ for $U/t=32$, $16$, and $8$, respectively.
We can see that $E_{min}$ decreases as one increases $S_{z}$ from
zero, i.e., it
costs less to dope a hole in the high-$S_{z}$ states than in the lower
$S_{z}$ states. So the doping of holes suppresses the gap and may stabilize
the high $S_{z}$ state. When the decrease of the hole energy becomes larger
than the energy difference between the ferromagnetic and the
antiferromagnetic configurations in the undoped case, the system goes through
a phase transition from the antiferromagnetism to the Nagaoka ferromagnetism.
In the large U limit, $\mu$ and $\Delta$ can be
calculated analytically by Taylor expanding Eqs. (\ref{gap}) and
(\ref{chem}),
\begin{equation}
\mu =  (-1/2 + N_{d}/N)U + O(t^{2}/U),
\label{chem2}
\end{equation}
\begin{equation}
\Delta =  [\frac{N_{d}}{N}(1-\frac{N_{d}}{N})]^{1/2}U + O(t^{2}/U).
\label{gap3}
\end{equation}
Here $N_{d}$ and $N$ is the number of doublons and lattice sites,
respectively, $S_{z} = N/2 - N_{d}$. The $S_{z} \neq 0$ case can again be
divided into two cases due to the constraint Eq. (\ref{cons}). First,
when $4t \geq (1/2-N_{d}/N)U$, from Eqs. (\ref{opti}) and
(\ref{chem2}) we get $\eta = at^{2}/U$ in
the large U limit, here $a$ is a number of order of 1. The minimum hole
energy
\begin{equation}
E_{min} = \Delta - U/2 + O(\frac{t^{4}}{U^{3}}).
\end{equation}
As $S_{z}$ increases from zero, $\Delta$ decreases
 since the number of the doublon-hole pairs
is equal to $(N/2-S_{z})$.
 Therefore $E_{min}$ decrease as $S_{z}$ increases from zero. Second,
when $4t < (1/2-N_{d}/N)U$, the optimum $(\epsilon_{k}-\mu)$ one can get
is $\eta = (1/2-N_{d}/N)U - 4t$ due to the constraint Eq. (\ref{cons}), and
\begin{equation}
E_{min} = \sqrt{\eta^{2} + \Delta^{2}} - U/2
\end{equation}
in the large U limit. Again, $E_{min}$ can be written as a Taylor series,
\begin{equation}
E_{min}(S_{z}) = -4t(1-\frac{2N_{d}}{N}) + O(t^{2}/U).
\end{equation}

Finally, we have also calculated the energy of two doped holes,
\begin{equation}
E_{2}=<\psi_{2}|H|\psi_{2}> - <\psi_{0}|H|\psi_{0}>,
\end{equation}
using the variational wavefunction,
\begin{equation}
|\psi_{2}> = \sum_{k,q} A_{k,q}
\gamma_{k}^{\dagger}\gamma_{q}^{\dagger}|\psi_{0}>.
\label{dope2}
\end{equation}
Here $A_{k,q}$ is determined by minimizing $E_{2}$. The quantity
$E_{2}-2E_{1}$ gives information about the interaction between
two holes. We find
\begin{equation}
E_{2}-2E_{1} = \frac{4U}{N} \sum_{k_{1},k_{2},k_{3}}
v_{k_{1}}v_{k_{2}}u_{k_{3}}u_{k_{2}+k_{3}-k_{1}}
A_{k_{1},k_{2}+k_{3}-k_{1}}A_{k_{2},k_{3}}^{*},
\end{equation}
which goes to zero in the thermodynamic
limit. This is because we did not allow the gap function $\Delta$ to
fluctuate in our calculation. It is believed that the fluctuation of the
antiferromagnetic
order parameter will result in an attractive interaction between holes, and
this attractive interaction favors a $d_{x^{2}-y^{2}}$ type of
pairing.\cite{pines}

To summarize the results of Sec. IV, we have extended the  pairing
wavefunction for the repulsive Hubbard model to the case of less than
half-filling  by using the Bogoliubov quasi-particle operators. We use
this variational wavefunction to calculate the energy of a doped hole. We
find that in the large U limit, the doped hole has a bandwidth of order of
J rather than t at $S_{z}=0$.\cite{brin} Here $J=4t^{2}/U$ is the magnetic
interaction
strength. The hole energy decreases as one increases $S_{z}$. The competition
between this effect and the Heisenberg spin interaction which favors
antiferromagnetism determines the transition between the antiferromagnetic
state and the Nagaoka ferromagnetic state. Our variational wavefunction
has the advantage of capable of representing a series of states in between
these two limits in a single, relatively simple form.

It also possible to view the system in the limit of high spin near
half filling as a low density, two component system.  Field theory methods
can be applied in this limit.  This approach lies outside the variational
method of this paper, but will be described in a future publication.

\section{Conclusions}

In conclusion, we have studied the metal-insulator transition and
magnetic ordering in half-filled Hubbard model using a particle-hole
mapping. We show that the 2D Hubbard model on any
lattice is an insulator at
half-filling for any finite U in the (Nagaoka) high spin region. The physical
picture of this metal-insulator transition
as a function of spin is quite
simple from the point of view of binding and unbinding of the hole-doublon
pairs. At high spin $S_{z}$, the low energy
excitations are neutral, consisting of the motion of hole-doublon pairs.
 In this region, the
dimensionality plays a very important role in determining whether the system
is a metal or a Mott insulator. For $d>2$, there exists a critical $U_{b}$,
when
and only when $U>U_{b}$, doublons and holes form charge neutral bound states
and the system is a Mott insulator in the high spin region; while for
$d\leq 2$, holes and doublons always form bound states and the system is an
insulator for any finite U in the high spin region. As one increases the
number of bound states, i.e., lowers $S_{z}$, the binding costs some extra
kinetic energy  if the Fermi surface is not
nested. When the cost of kinetic energy
exceeds the binding energy, a binding-unbinding transition takes place. The
excitations are charged, and the system is a metal. If the Fermi surface is
nested, however, this binding costs no extra kinetic energy, and the
system is always an insulator for any finite U at any spin $S_{z}$.
We have also
discussed the generalization of the pairing wavefunction (\ref{bcs}) to the
hole doped case using Bogoliubov quasiparticle operators.   We find that the
dispersion of the doped hole is significantly changed due to the presence of
the interaction. In the large U limit, the doped hole has a bandwidth of
order of J rather than t at $S_{z}=0$.

\acknowledgements

It is a pleasure to thank Elbio Dagotto, Richard Ferrell, Steve Girvin,
Adriana Moreo, Mohit Randeria, and Nandini Trivedi
for valuable discussions. We would also like to thank the hospitality of the
Aspen Center for Physics
where part of the work was done. This work was supported by the NSF
through Grant No. DMR 9214739, and by the Electric Power Research
Institute. We also acknowledge support from San Diego Supercomputer
Center.

\begin{figure}
\caption[]{Critical interaction $U_{b}$ as a function of the interlayer
hopping constant $t^{\prime}$ in a 3D bipartite lattice. When $U>U_{b}$, a
pair of hole and
doublon forms charge neutral bound state and Hubbard model is an insulator
at the high spin region.  }\label{inter}
\end{figure}

\begin{figure}
\caption[]{(a). Fermi surfaces of doublons (solid lines) and holes
(dashed lines)
of the 2D Hubbard model on a square lattice with
the nearest neighbor hopping only at various $S_{z}$.
(b). Fermi surfaces of doublons (solid lines) and holes (dashed lines)
of the 2D Hubbard model on a square lattice with
the nearest and the next nearest neighbor hopping at $S_{z}=0.4N$, $0.3N$, and
$0.2N$. The parameter is $t_{2}=0.25t$. }
\label{fs}
\end{figure}

\begin{figure}
\caption[]{Critical interaction $U_{c}$ of the 2D Hubbard model on a square
lattice with NN and NNN hopping as a function of $S_{z}$.
When $U>U_{c}$, holes and doublons  form charge neutral bound states and the
system is an insulator. The parameter is the same as in Fig. 2(b). }
\label{uc}
\end{figure}

\begin{figure}
\caption[]{(a) Fermi surfaces of doublons (solid lines) and holes (dashed
lines)
of the Hubbard model on a 2D triangular lattice with
the nearest neighbor hopping only at various $S_{z}$.
 (b).  Critical interaction $U_{c}$ of the Hubbard model on a triangular
lattice with NN hopping as a function of $S_{z}$.
When $U>U_{c}$, holes and doublons  form charge neutral bound states and the
system is an insulator. }
\label{tri}
\end{figure}

\begin{figure}
\caption[]{Shows the calculated (a) minimum hole energy, and (b) the
bandwidth of a single doped hole  of the 2D
Hubbard model on a square lattice with the nearest neighbor hopping as a
function of $U/t$ at $S_{z}=0$ and one-hole away from half-filling. }
\label{enh}
\end{figure}

\begin{figure}
\caption[]{Shows the minimum hole energy of 2D Hubbard model on a square
lattice as a function of $S_{z}$ for $U=32t$
(solid), $16t$ (dashed), and $8t$ (dotted), respectively. }
\label{enh2}
\end{figure}


\begin{references}
\bibitem{Hubb} J. Hubbard, Proc. Roy. Soc. London A {\bf 276}, 238
(1963); {\bf 281}, 401 (1964).
\bibitem{frad} E. Fradkin, {\em Field Theories of
condensed matter systems}, (Addison-Wesley, Redwood City, California,
1991).
\bibitem{mott} N. F. Mott, {\em Metal Insulator Transitions}, (Taylor
and Francis, London, 1974).
\bibitem{hirs} J. E. Hirsch, Phys. Rev. B {\bf 31}, 4403 (1985); H. Q. Lin
and J. E. Hirsch, Phys. Rev. B {\bf 35}, 3359 (1987).
\bibitem{nag} Y. Nagaoka, Phys. Rev. {\bf 147}, 392 (1966).
\bibitem{ande} P. W. Anderson, Solid State Phys. {\bf 14}, 99 (1963).
\bibitem{azr} P. W. Anderson, Science {\bf 235}, 1196 (1987); F. C.
Zhang and T. M. Rice, Phys. Rev. B {\bf 37}, 3759 (1988).
\bibitem{keme} G. Kemeny and L. G. Caron, Rev. Mod. Phys. {\bf 40},
790 (1968); Y. Nagaoka, Prog. Theor. Phys. {\bf 52}, 1716 (1974);
V. J. Emery, Phys. Rev. B {\bf 14}, 2989 (1976);
R. R. P. Singh and R. T. Scalettar, Phys. Rev. Lett. {\bf
66}, 3203 (1991).
\bibitem{kris} H. R. Krishnamurthy, C. Jayaprakash, S. Sarker, and W.
Wenzel, Phys. Rev. Lett. {\bf 64}, 950 (1990).
\bibitem{li} Q. P. Li and R. Joynt, Phys. Rev. B {\bf 47}, 3979 (1993).
\bibitem{leg} A. J. Leggett, in {\em Modern Trends in the Theory of
Condensed Matter}, edited by A. Pekalski and R. Przystawa (Springer,
Berlin, 1980).
\bibitem{nozi} P. Nozieres and S. Schmitt-Rink, J. Low Temp. Phys.
{\bf 59}, 195 (1985).
\bibitem{rand} M. Randeria, J.-M. Duan, and L.-Y. Shieh, Phys. Rev.
Lett. {\bf 62}, 981 (1989); K. Miyake, Prog. Theor. Phys. {\bf 69}, 1794
(1983).
\bibitem{schr:sc} J. R. Schrieffer, {\em Theory of Superconductivity},
(Benjamin/Cummings, Reading, MA, 1971).
\bibitem{kohn} W. Kohn, Phys. Rev. {\bf 133}, A171 (1964).
\bibitem{mill} A. J. Millis and S. N. Coppersmith, Phys. Rev. B {\bf
43}, 13770 (1991).
\bibitem{fuld} P. Fulde and R. A. Ferrell, Phys. Rev. {\bf 135}, A550
(1964); C. N. Yang, Phys. Rev. Lett. {\bf 63}, 2144 (1989).
\bibitem{lark} A. I. Larkin and Yu. N. Ovchinnikov, Zh. Eksp. Teor.
Fiz. {\bf 47}, 1136 (1964). [Sov. Phys. JETP {\bf 20}, 762 (1965).]
\bibitem{dege:sc} P. G. de Gennes, {\em Superconductivity of Metals
and Alloys}, (Addison-Wesley, Redwood City, CA, 1989) Chapter 4.
\bibitem{sark} S. Sarker, C. Jayaprakash, H. R. Krishnamurthy, and W.
Wenzel, Phys. Rev. B {\bf 43}, 8775 (1991);   C. Jayaprakash,
H. R. Krishnamurthy, S. Sarker, and W. Wenzel, Europhys. Lett. {\bf
15}, 625 (1991).
\bibitem{bogo} N. N. Bogoliubov, {\em Nuovo Cimento} {\bf 7}, 794
(1958); J. G. Valatin, {\em Nuovo Cimento} {\bf 7}, 843 (1958).
\bibitem{dag} E. Dagotto, A. Moreo, R. Joynt, S. Bacci, and
E. Gagliano, Phys. Rev. B {\bf 41}, 2585 (1990).
\bibitem{von} K. von Szczepanski, P. Horsch, W. Stephan, and
M. Ziegler, Phys. Rev. B {\bf 41}, 2017 (1990).
\bibitem{trug} S. Trugman, Phys. Rev. B {\bf 41}, 892 (1990); {\bf 37}, 1597
(1988).
\bibitem{poil} D. Poilblanc and E. Dagotto, Phys. Rev. B {\bf 44}, 466
(1991).
\bibitem{chub} A. V. Chubukov and D. M. Frenkel, Phys. Rev. B {\bf 46},
11884 (1992).
\bibitem{dag2} E. Dagotto, preprint.
\bibitem{pines} P. Monthoux, A. V. Balatsky, and D. Pines, Phys. Rev.
Lett. {\bf 67}, 3448 (1991); P. Monthoux and D. Pines, Phys. Rev.
Lett. {\bf 69}, 961 (1992).
\bibitem{brin} W. F. Brinkman and T. M. Rice, Phys. Rev. B {\bf 2}, 1324
(1970).
\end{references}
\end{document}